# High-energy femtosecond Yb:CaF$_2$ laser for efficient THz pulse generation in lithium niobate

C. Vicario,[1,*] B. Monoszlai,[1,2] Cs. Lombosi,[2] A. Mareczko,[3] A. Courjaud,[3] J. A. Fülöp[4,5] and C. P. Hauri[1,6]

[1]*Paul Scherrer Institute, SwissFEL, 5232 Villigen-PSI, Switzerland*
[2] *Institute of Physics, University of Pécs, Ifjúság ú. 6, 7624 Pécs, Hungary*
[3]*Amplitude Systèmes, Pessac, France*
[4] *MTA-PTE High-Field Terahertz Research Group, Ifjúság ú. 6, 7624 Pécs, Hungary*
[5]*ELI-Hu Nkft., Dugonics tér 13, 6720 Szeged, Hungary*
[6]*Ecole Polytechnique Federale de Lausanne, 1015 Lausanne, Switzerland*

We present a study on THz generation in lithium niobate pumped by a powerful and versatile Yb:CaF$_2$ laser. The unique laser system delivers transform-limited pulses of variable duration (0.38-0.65 ps) with pulse energy of up to 15 mJ at a center wavelength of 1030 nm. From theoretical investigations it is expected that those laser parameters are ideally suited for efficient THz generation. Here we present experimental results on both the conversion efficiency and the THz spectral shape for variable pump pulse durations and for different crystal temperatures down to 25 K. We experimentally verify the optimum pump parameters for most efficient and broadband THz generation.

High-energy ultrashort terahertz (THz) pulses in the frequency range of 0.1-10 THz (photon energy between 0.4 and 40 meV) allow direct access to several resonant and nonresonant excitations and collective modes, such as phonons, magnons and electromagnons [1]. The key advantage of terahertz pulses with respect to the more conventional visible and near-infrared laser is that these modes can be excited without significant heat deposition. Intense THz pulses opens new applications such as non-resonant excitations of magnetization dynamics [2], THz-assisted attosecond pulse generation [3], initiation of catalytic reactions [4], and optical THz undulators [5], which would benefit from field strength higher than currently provided from table-top sources.

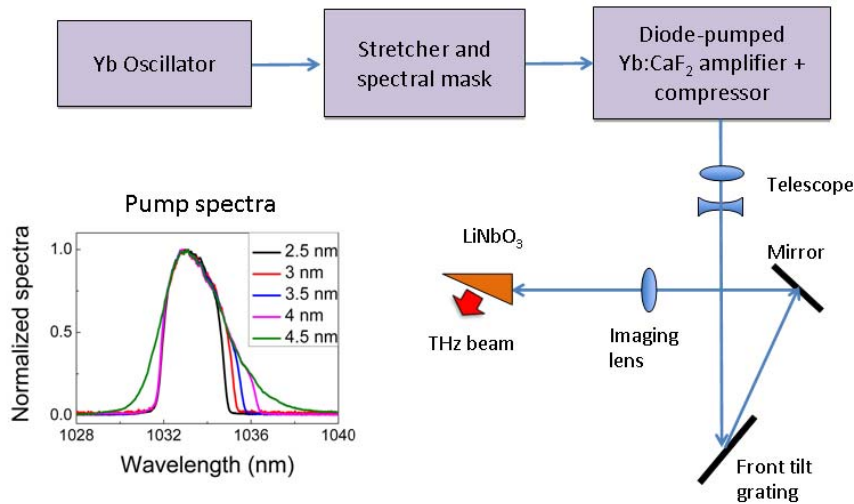

Fig. 1. Experimental setup: Yb:CaF2 CPA laser delivers sub-ps, 15 mJ pump pulses; after down-collimation and pulse front tilt the beam is directed to LN for THz generation. Shown in the inset are the laser spectra produced by different spectral cut.

Optical rectification (OR) of intense laser pulses in nonlinear crystals is one of the most promising approaches to generate high-energy terahertz pulses in the range of 0.1-10 THz with table-top systems [6-11]. While organic salt crystals offer efficient generation between 1-10 THz, the inorganic LiNbO$_3$ (LN) demonstrated potential for high-energy pulse generation below 1 THz. By using LN at room temperature (RT) and pump pulses with a typical duration of about 100 fs pump-to-THz conversion efficiencies up to 10$^{-3}$ could be demonstrated [9]. It was proposed theoretically that OR in LN could be significantly improved by optimizing the Fourier-limited (FL) pump pulse duration and cooling LN to cryogenic temperature (CT) to reduce its THz absorption [12]. It was predicted that efficient laser sources with a central wavelength around 1 μm delivering transform-limited pulses of about 600 fs are ideally suited for driving OR in LN. Recent experimental studies confirmed these predictions. The highest so far THz pulse energy (125 μJ) from OR could be reached with longer (though non-optimal) pump pulses [10]. The highest conversion efficiency (3.8%) was demonstrated by cryogenic cooling of congruent LN and using close-to-optimal pump pulse duration [11].

In this Letter we present a versatile, powerful sub-picosecond Yb:CaF$_2$ laser system which is employed for investigation of the optimal pump parameters for efficient and broadband THz generation. The laser optimization study has been performed in stoichiometric LN, which is used at room temperature (RT) as well as at cryogenic temperature (CT, 25 K). In order to find the maximum conversion efficiency we investigated OR with variable, but at all time FL pulse durations at both temperatures and report conversion efficiencies as well as the corresponding THz spectra. To our best knowledge this is the first systematic study on THz spectra and efficiency as function of the FL pump pulse duration and crystal temperature. While emphasis has been put solely on enhanced efficiency in the past, the potential changes of the THz spectral shape have not been examined.

The experimental setup is sketched in Fig. 1. The laser system is based on chirped-pulse amplification (CPA) and consists of a compact ytterbium oscillator, a grating-based stretcher-compressor assembly and a regenerative Yb:CaF$_2$ amplifier operated at room temperature. The broadband oscillator running at a repetition rate of 54 MHz provides 1-nJ pulses that are stretched to 500 ps prior to injection into the regenerative amplifier. The 4-mm long amplifier gain medium is pumped from the two opposing sides by two 80-W CW diode modules emitting at 981 nm. This single-stage compact amplifier provides up to 30 mJ uncompressed beam at a repetition rate of 10 Hz.

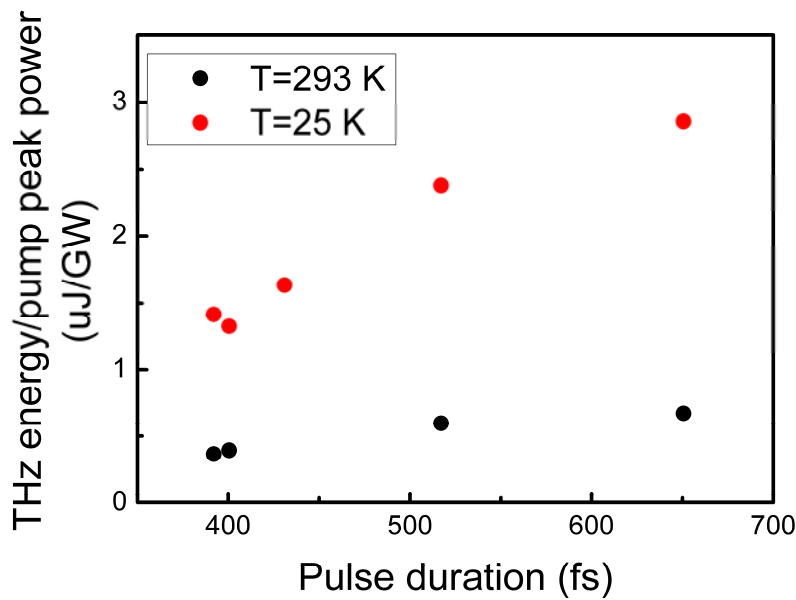

Fig. 2. THz pulse energy normalized to the pump peak power as function of the laser pulse duration at T= 293 K and T=25 K.

As particular feature the laser design allows seeding in narrowband as well as broadband mode without risk of damage while maintaining the 15 mJ pulse energy after compression. In the inset of Fig. 1 different spectral widths are shown, recorded after the compressor, resulting in near-FL pulses of durations between 380 fs and 650 fs.

For THz generation the intense laser beam is down-collimated by a 3:1 telescope and directed to the LN. In order to achieve the phase matching with the THz radiation, the pump pulse front is tilted using a reflective diffraction grating [10]. The surface of the diffractive optics is then imaged on the input surface of the LN crystal using a spherical lens of 250 mm focal length. Pulse energy up to 12 mJ over a 5 mm beam diameter is available to pump the LN crystal.

The OR is carried out in stoichiometric LiNbO$_3$ doped with 0.6% MgO. The crystal input surface is 8 by 16 mm and the LN is cut at 63°. The crystal is installed in a cryostat and it can be cooled down to 25 K. The THz pulse is then characterized in energy by means of a calibrated pyroelectric detector. Spectral measurements are performed with an in air THz Michelson interferometer (Fourier-transform spectroscopy).

In Fig. 2 the THz energy normalized to the pump peak power is plotted as function of the laser pulse duration. As expected from theory [12] the experimental investigations unravel a clear dependence of the conversion efficiency on the pulse duration. Maximum pump-to-THz conversion efficiency of 0.36% is reached for the 650-fs pulse, the longest pulse available from our laser system. For this duration the efficiency is twice as high as that measured for the shortest pump pulse under otherwise equal conditions. Similar behavior is observed both at RT and at 25 K. The measurements indicate furthermore that, for specific pulse duration, the conversion efficiency is indeed systematically higher for cryogenically cooled LN. While higher conversion efficiency is doubtlessly whished it needs to be clarified how the THz spectrum is affected by the pulse duration. We therefore experimentally investigate the change of THz spectral output as function of the FL pump pulse duration and the LN temperature.

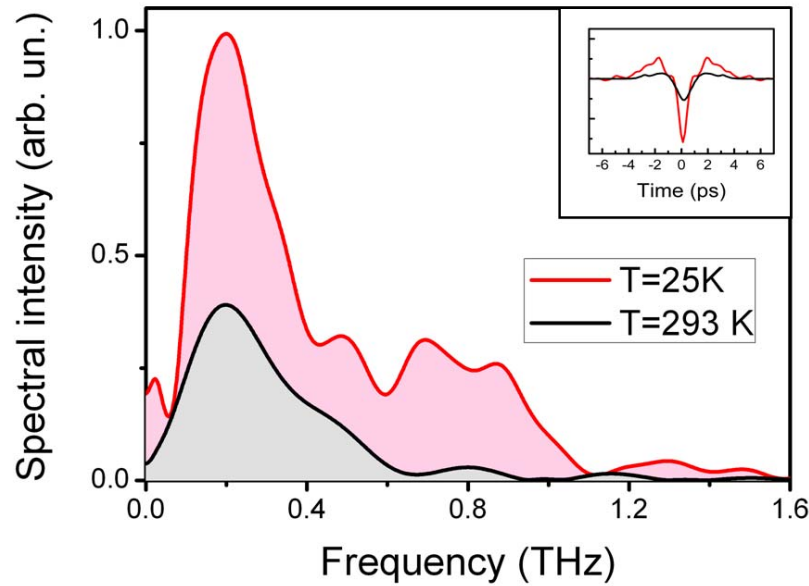

Fig. 3. Cryo-cooled LiNbO3 gives rise to higher frequencies and threefold higher energy at (red curve) low temperature in comparison with emission at (black curve) room temperature. THz spectra are reconstructed by Fourier-transformation of the interferograms shown in the inset.

Shown in Fig. 3 are THz spectra at (black curve) RT and at (red curve) CT. The spectral intensity is retrieved by Fourier-transformation of the interferograms reported in the inset of Fig. 3. The spectra at both the temperatures are generated using the shortest available pump pulse (380 fs FWHM). Obviously the THz radiation covers a significantly broader range at CT shifting the drop-off frequency from 0.6 THz (RT) to 1.1 THz (CT). The occurrence of broader spectrum at 25 K is predicted by theory and it is due to the dramatic reduction in absorption for higher frequencies in comparison to RT. Some of the spectral modulations are associated to water absorption peaks (0.55 THz; 0.75 THz) and are in agreement with the data reported in the literature [13]. The corresponding interferograms indicate THz pulse duration slightly longer than 1 ps (FWHM). The results show that with LN at cryogenic temperature it is possible to fully cover the spectral region between 0.1 to 1 THz.

Reported in Fig. 4 are the normalized THz spectra at (a) RT and (b) CT for different pump pulse durations. Our experiment shows clearly that a shorter pump pulse gives rise to higher-frequency components. The enhancement of higher THz frequencies is more evident for cooled LN (Fig. 3b) because the high-frequency THz absorption is significantly reduced at CT. Our investigations show that broader THz spectra require shorter pump pulses, which inherently goes along with lower conversion efficiency, as shown in Fig. 2.

At CT the $LiNbO_3$ produces three times larger integrated signal as compared to RT. The maximum conversion efficiency achieved in our experiment is 0.12 % at RT and 0.36% at 25 K. Under the latter conditions the largest THz pulse energy reached 45 µJ. The measured RT conversion efficiencies are similar to the values measured previously at longer pulses [10], and are below the theoretical predictions [12]. The results are in contrast to recently reported experimental values [11]. The reason for this discrepancy needs further investigation, which is out of the scope of the present study, and may also be connected to impurities in the LN [12] or to the different typology of the crystal. The high pulse energy could potentially result in peak electric field of 2 MV/cm. This value is calculated assuming central frequency equal to the spectrum centroid at 0.4 THz and a focusing system with numerical aperture of 1.

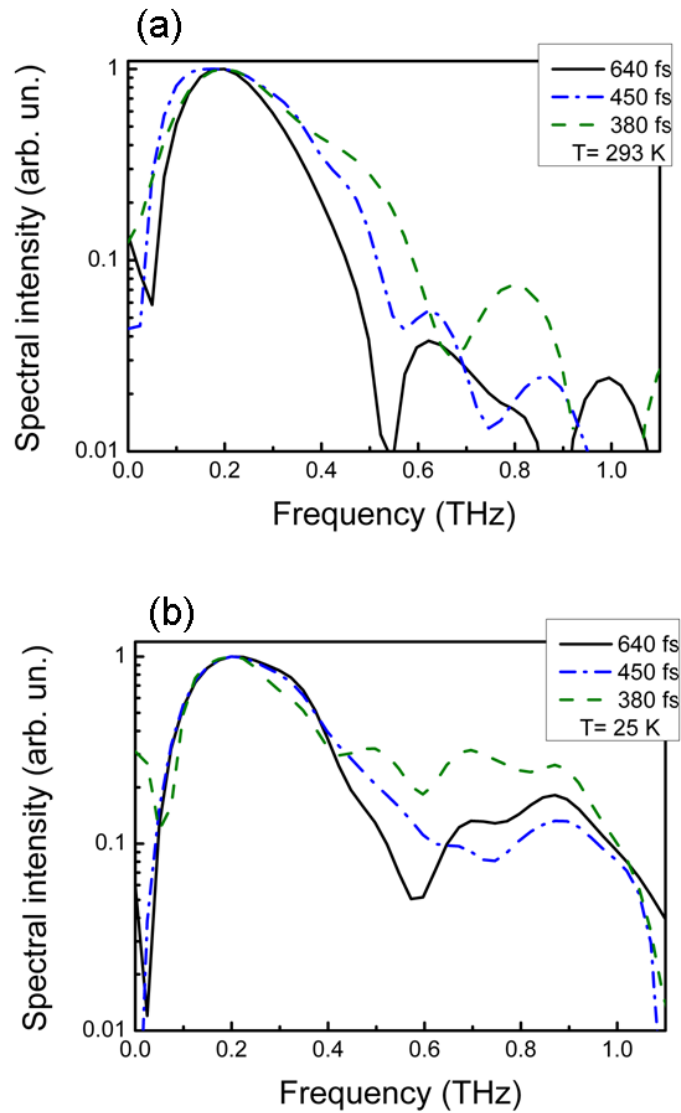

Fig. 4. THz spectral intensity for different pump pulse durations at RT and at T=25 K.

In conclusion, we explored optimum conditions for intense THz pulse generation in LiNbO$_3$ driven by a multi-mJ compact Yb:CaF$_2$ laser. The efficiency of OR and emitted THz spectra are studied for various LiNbO$_3$ temperatures and pump pulse durations. While broadest THz spectra (0.1-1.2 THz) are achieved by using the shortest available pump pulses, longer pulses were beneficial for reaching highest energy conversion efficiency (0.36%) and pulse energies (45 uJ) at the cost of THz bandwidth. For all pulse durations cryogenic cooling of the LN systematically leads to higher conversion efficiencies and higher frequency components. Our experimental results indicate that the versatile Yb:CaF$_2$ laser could be an ideal tool for tailoring the THz output to the need of the specific THz application.

This work was partially supported by the Swiss National Science Foundation under grant PP00P2_128493, SCIEX-NMS[ch] (grant no. 12.159) and NSF (grant no. 51NF40-144615), as well as by the Hungarian Scientific Research Fund (OTKA) grant no. 101846, the SROP-4.2.1.B-10/2/KONV-2010-0002 and hELIos ELI_09-01-2010-0013 projects. The authors gratefully acknowledge Fastlite for the loan of equipment, and J. Hebling for helpful discussion. CPH acknowledges also association to the Centre of Competence in Research (NCCR-MUST).